\documentclass[a4paper,11pt]{article}
\pdfoutput=1 
\usepackage{jcappub} 
\usepackage{amsfonts}
\usepackage[T1]{fontenc} 
\usepackage{ulem}
\usepackage{amsmath}
\usepackage{dsfont}
\usepackage{epstopdf}
\usepackage{ulem}
\usepackage{graphicx}
\usepackage{bbold}
\usepackage{ esint }

\usepackage{textcomp} 
\usepackage{wasysym} 
\usepackage{bm}
\usepackage{esint}
\usepackage{xcolor}
\usepackage{cancel}
\usepackage{comment}
\usepackage{hyperref}
\hypersetup{
    colorlinks=true,
    linkcolor=red,
    citecolor=blue,
} 
\newcommand{\change}[1]{#1}

\newcommand{\W}[0]{{\rm{W}}}

\renewcommand{\d}[0]{{\rm{fl}}}

\newcommand{\mO}{{\mathcal O}}
\newcommand{\del}[0]{\partial }

\newcommand{\vol}[2]{\hspace{-0.8mm}\mbox{$\text{d}^{\hspace{-0.0mm}#1}$}\hspace{-0.2mm}#2\hspace{0.8mm}\ }
\renewcommand{\v}[1]{\bm{#1} }
\newcommand{\vv}[0]{\bm{v} }
\newcommand{\vw}[0]{\bm{w} }
\newcommand{\vJ}[0]{\bm{J} }
\newcommand{\vx}[0]{\bm{x} }

\newcommand{\vp}[0]{\bm{p} }
\newcommand{\vq}[0]{\bm{q} }
\newcommand{\vnabla}[0]{\bm{\nabla} }

\title{Finding closure: approximating Vlasov-Poisson using finitely generated cumulants}

\author[a,b]{Cora Uhlemann}
\affiliation[a]{Centre for Theoretical Cosmology, DAMTP, University of Cambridge, \\
CB3 0WA Cambridge, United Kingdom}
\affiliation[b]{Fitzwilliam College, University of Cambridge, CB3 0DG Cambridge, United Kingdom}

\emailAdd{c.uhlemann@dampt.cam.ac.uk}

\abstract{Since dark matter almost exclusively interacts gravitationally, the phase-space dynamics is described by the Vlasov-Poisson equation. A key characteristic is its infinite cumulant hierarchy, a tower of coupled evolution equations for the cumulants of the phase-space distribution. While on large scales the matter distribution is well described as a fluid and the hierarchy can be truncated, smaller scales are in the multi-stream regime in which all higher-order cumulants are sourced through nonlinear gravitational collapse. This regime is crucial for the formation of bound structures and the emergence of characteristic properties such as their density profiles.
We present a novel closure strategy for the cumulant hierarchy that is inspired by finitely generated cumulants and hence beyond truncation. This constitutes a {\it constructive} approach for reducing nonlinear phase-space dynamics of Vlasov-Poisson to a  closed system of equations in position space.

Using this idea, we derive Schr\"odinger-Poisson as approximate quantal method for solving classical dynamics of Vlasov-Poisson with cold initial conditions. Our deduction complements the common reverse inference of the Schr\"odinger-Vlasov relation 
using a semi-classical limit of quantum mechanics and provides a clearer picture of the correspondence between classical and quantum dynamics. Our framework outlines an essential first step towards constructing approximate methods for Vlasov-like systems in cosmology and plasma physics with different initial conditions and potentials.
}

\begin{document}

\maketitle

\section{Introduction}

A multitude of cosmological probes \cite{Roos12} have established that the dominant matter component is of unknown dark origin, exhibiting only very small, if any, non-gravitational interactions. Dark matter is indispensable for our understanding of cosmic structure formation, as it is able to cluster early due to the absence of forces opposing gravity, as well as providing the environment for galaxy formation \cite{Wechsler18,Desjacques18}. This property is highly beneficial from a theoretical point of view, as it admits a two-step approach: first
solving purely gravitational collisionless dynamics for the dark matter component dominating the large scales, and secondly tackling the more complicated interplay between gravitational and baryonic effects that mostly affects smaller scales. Splitting the problem allowed for major advances in the theoretical description \cite{Zeldovich70,Peebles80,Bernardeau02,Baumann12} and numerical modelling \cite{Springel05,MICE,HR4} of cosmic large scale structure. To take full advantage of observational data from massive galaxy surveys \citep{DES,KIDS,Euclid,LSST,WFIRST}, such as testing the cosmological standard model and probing fundamental physics, we need to push our predictions to higher precision and smaller scales.

 There are three long-standing goals in cosmic structure formation that require further theoretical progress to understand dark matter dynamics in the nonlinear regime:
\begin{itemize}
\item establish a consensus between different solution strategies (based on discrete particles or continuous fields, treated numerically or semi-analytically) and their associated artefacts, 
in particular on small scales where bound structures form;
\item incorporate physical effects of massive neutrinos or non-standard dark matter properties, 
in order to probe fundamental physics through cosmology;
\item understand striking universalities in the outcome of nonlinear gravitational collapse, 
such as properties of the cosmic web as a whole or individual bound structures.
\end{itemize}

N-body simulations have been established as the state-of-the-art benchmark for testing theoretical models and extracting fitting functions. The volume and the number of particles that can be simulated are steadily growing with the computational power of supercomputers. However, even the largest cosmological simulations to date with more than a trillion particles \cite{Potter17} simulate particles with masses of the order of a billion solar masses and hence effective mass points representing huge conglomerates of dark matter rather than actual particles. Numerical artefacts intrinsic to N-body methods, such as sparse sampling and unphysical two-body effects, can pose a challenge for correctly resolving the dynamics of the coherent dark matter field down to small scales where one is sensitive to the particle properties of dark matter. This is why there is increasing activity in tackling the full phase-space dynamics numerically \cite{Colombi08,Shandarin12,Yoshikawa13,Colombi14,Hahn16,Sousbie16} or recreating the phase-space structure from $N$-body simulations \cite{Abel12}.
An alternative is to use approximate field-based methods inspired by the quantum-classical correspondence which allows to reduce dynamics from phase-space to position space \cite{Uhlemann14,Kopp17,Garni18,Mocz18}. Potential generalisations of this correspondence inspired from quantum field theory are discussed in \cite{Prokopec17,Friedrich18}.

The quantum-inspired approach has gained considerable interest recently due to its connection to alternative types of dark matter, such as ultralight scalar fields or axions, that behave like standard dark matter on large scales but leave characteristic imprints on smaller scales \cite{Schive14,Niemeyer16} and in other astrophysical probes \cite{Hui17,Marsh16}. Aside from this, the presence of massive neutrinos modifies the clustering on small scales and is one of the prime targets for probing fundamental physics through cosmology. It proves challenging to include massive neutrinos with large thermal velocities in N-body simulations, because their momentum distribution needs to be sampled sufficiently accurate \cite{Banerjee18,Brandbyge18} requiring an excessive number of particles. This makes it desirable to find a field-based technique that can be applied to warm initial conditions with a given momentum distribution.

Cosmological simulations have revealed striking universal characteristics of the cosmic web as a whole \cite{Gott86,Kofman88,Shandarin89,vandeWeygaert09,Origami12,Codis18} and individual structures such as the density profiles of bound dark matter halos \cite{NFW97,Ludlow13,Angulo17} and voids \cite{Hamaus14,Nadathur15,Cautun16}. Those findings suggest that there is an underlying principle and an approximate method to reproduce those features without running billion particle simulations. Beyond this somewhat academic motivation, it is also of great practical interest to develop models for the formation of dark matter halos. The halo model has emerged as a key tool to describe the outcome of dark matter clustering  \cite{PressSchechter74,Peacock00,Seljak00,MoWhite96,ShethTormen99,Cooray02} by associating all dark matter with halos as bound structures. Then, the statistical properties of the large-scale density can be obtained from a prediction of the number and spatial distribution of the halos, as well as from the distribution of matter within each halo.

Achieving these goals requires to understand the so-called multi-stream regime, in which dark matter cannot be treated as a perfect fluid described by just density and velocity. This regime emerges naturally when dark matter particles collapse onto an overdensity where they shell-cross because of their collisionless nature. So far, the multi-stream regime is mostly the domain of numerical simulations, while only few analytical approaches for the treatment of shell-crossing \cite{Taruya17,Rampf17,Saga18} or its long-term limits \cite{FillmoreGoldreich84,Pietroni18} are being developed. A further complication arises from the fact that bound structures are formed through a whole series of crossings that successively increases the number of streams such that one needs a method that can dynamically describe the formation of new streams. Here, we will approach this problem from a theoretical angle that looks at the phase-space distribution functions and possible expansions in terms of cumulants.

The paper is structured as follows.
Section~\ref{sec:VlasovPoisson} presents the Vlasov-Poisson system as the equation of motion for the phase-space distribution of dark matter and derives the associated cumulant hierarchy of gravitational collapse. It briefly discusses the associated initial conditions and the qualitative phenomenology of the time evolution. 
Section~\ref{sec:FGC} introduces strategies for dealing with cumulant hierarchies by starting from the perfect fluid model and then introducing the concept of finitely generated cumulants for simple distribution functions and finally phase-space distributions.
Section~\ref{sec:SPder} presents a derivation of the Schr\"odinger-Poisson system from applying the idea of finitely generated cumulants to the Vlasov-Poisson system.
Section~\ref{sec:Conclusion} concludes, puts the results in a broader context of physics and indicates some possible ways forward.

\section{The Vlasov-Poisson hierarchy for cold dark matter}
\label{sec:VlasovPoisson}

\subsection{The Vlasov-Poisson equation describing cold dark matter}
On scales that are small compared to the Hubble radius (and hence the observable universe), in the weak field regime and for non-relativistic velocities, one can use the Newtonian limit rather than the full Einstein equations to describe the time evolution of structures within the universe \cite{CZ11,GW12,KUH14}. Furthermore, since we are interested in the dynamics of a collection of dark matter particles which are very abundant in the universe, collisional effects are completely negligible as they are suppressed by the total number of particles \cite{Gilbert68}. Hence, instead of having to solve the BBGKY (Bogoliubov-Born-Green-Kirkwood-Yvon) hierarchy for the series of $n$-point phase-space distributions $f_n(\{\vx_i,\vp_i\}_{i=1,\ldots,n})$, we just have to solve a collisionless equation for the one-particle phase-space distribution $f=f_1(\vx,\vp)$. We will consider the phase-space distribution in the continuum limit, where it is not a collection of peaks at the sites of all of the particles, but a coherent field encoding the probability of finding particles in a \change{phase-space} volume.

The Vlasov-Poisson system for the one-particle phase-space density $f(t,\vx,\vp)$ describes the time evolution of collisionless dark matter under gravitational interaction, in the absence of two-body interactions. It essentially follows from the conservation of phase-space $df/dt=0$ and can be rephrased in terms of the Poisson-bracket $\{H,f\}$ of the Hamiltonian $H$ of the system and the phase-space distribution $f$ which one can spell out in operator notation

\begin{subequations}
\label{VlasovPoissonEq}
\begin{align}
\label{VlasovEq}
\partial_t f&=   -\frac{\vp}{a^2 m}\cdot\vnabla_{\! \!  x} f + m \vnabla_{\! \!  x} V \cdot\vnabla_{\! \!  p} f 
 \\
 &=\left[ \frac{\vp^2}{2a^2m}+m V(\vx)\right] \left( \change{\overleftarrow{\vnabla}_{\! \!  x} \cdot \overrightarrow{\vnabla}_{\! \!  p}- \overleftarrow{\vnabla}_{\! \!  p} \cdot\overrightarrow{\vnabla}_{\! \!  x}}\right) f =\{H,f\}\,,
\end{align}
where derivative operators with arrows $\overleftarrow{\vnabla}$/$\overrightarrow{\vnabla}$ indicate whether they are acting on functions to their left or right, $V$ is the gravitational potential, which depends on the density that is an integral of the phase-space distribution according to the Poisson equation
\begin{align}
\label{PoissonEq}
\Delta V &= \frac{4\pi G\,\rho_0}{a}  \left(\int \vol{3}{p}\!\!f - 1\ \right) \,.
\end{align}
\end{subequations}
The Vlasov equation is a partial differential equation for the phase-space distribution function $f(t,\vx,\vp)$ involving cosmic time $t$ and 3+3-dimensional phase-space $(\vx,\vp)$ as variables. Our description here makes use of \textit{comoving} coordinates $\vx$ which are related to physical coordinates as $\v{r}=a(t)\vx$ where $a(t)$ is the scale factor that determines the time evolution of the background FRW universe. If one considers a non-expanding case, one can set $a=1$. The momentum $\vp=a^2 m d\vx/dt$ is conjugate to the comoving spatial variable $\vx$. 
The Poisson equation encodes the gravitational interaction and makes the Vlasov-Poisson equation a coupled nonlinear, partial, integro-differential equation.

\paragraph*{Initial conditions} 
The initial conditions are usually specified as cold (with negligible initial velocities) $f(t=t_0,\vx,\vp)=\rho(\vx)\delta_D(\vp)$ which makes solving the Vlasov-Poisson equation a problem of following an initially flat 3-dimensional phase-space sheet in the course of its time evolution in 6-dimensional phase-space.

\begin{figure}[h!]
\includegraphics[width=0.32\textwidth]{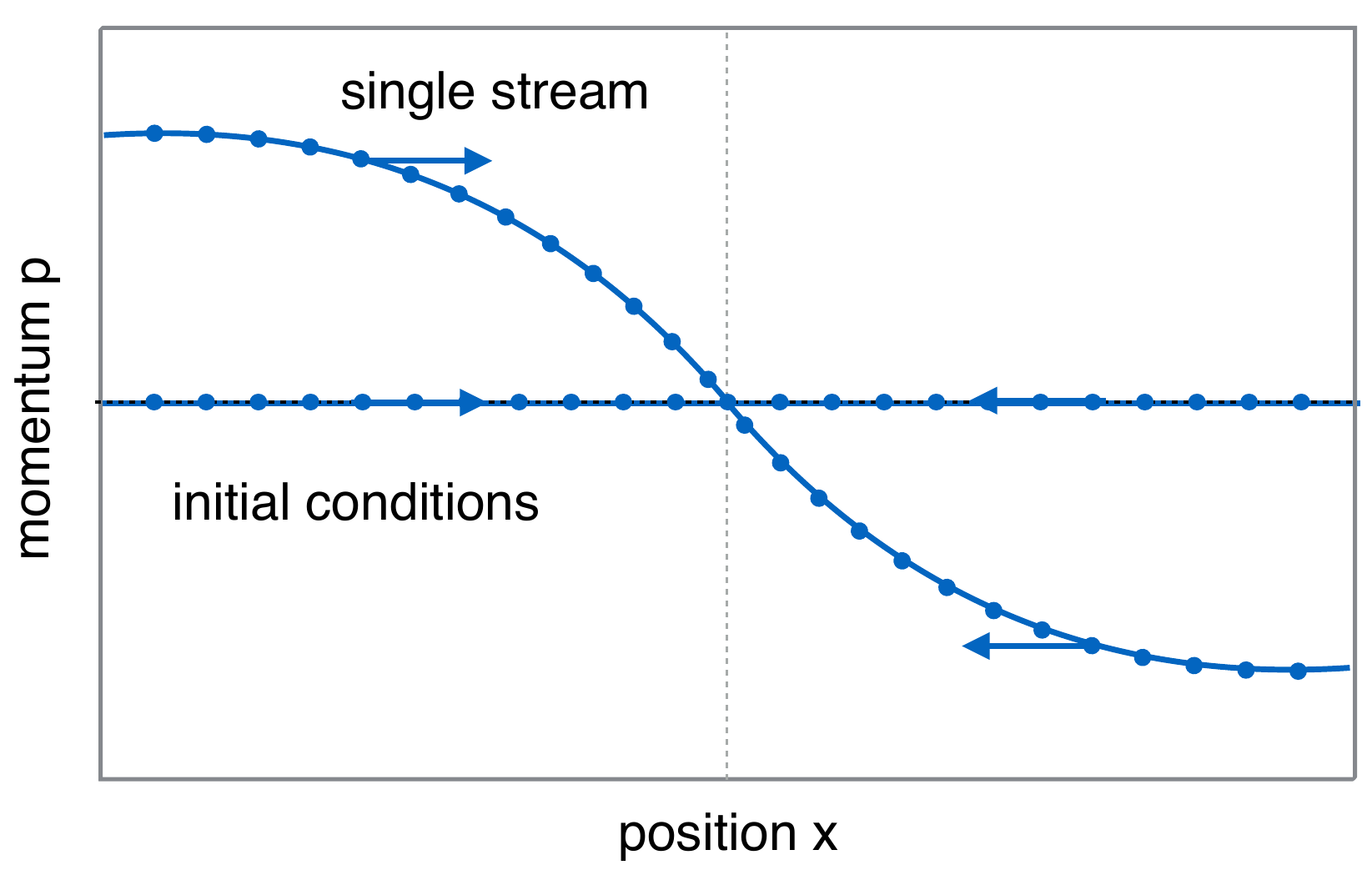} 
\includegraphics[width=0.32\textwidth]{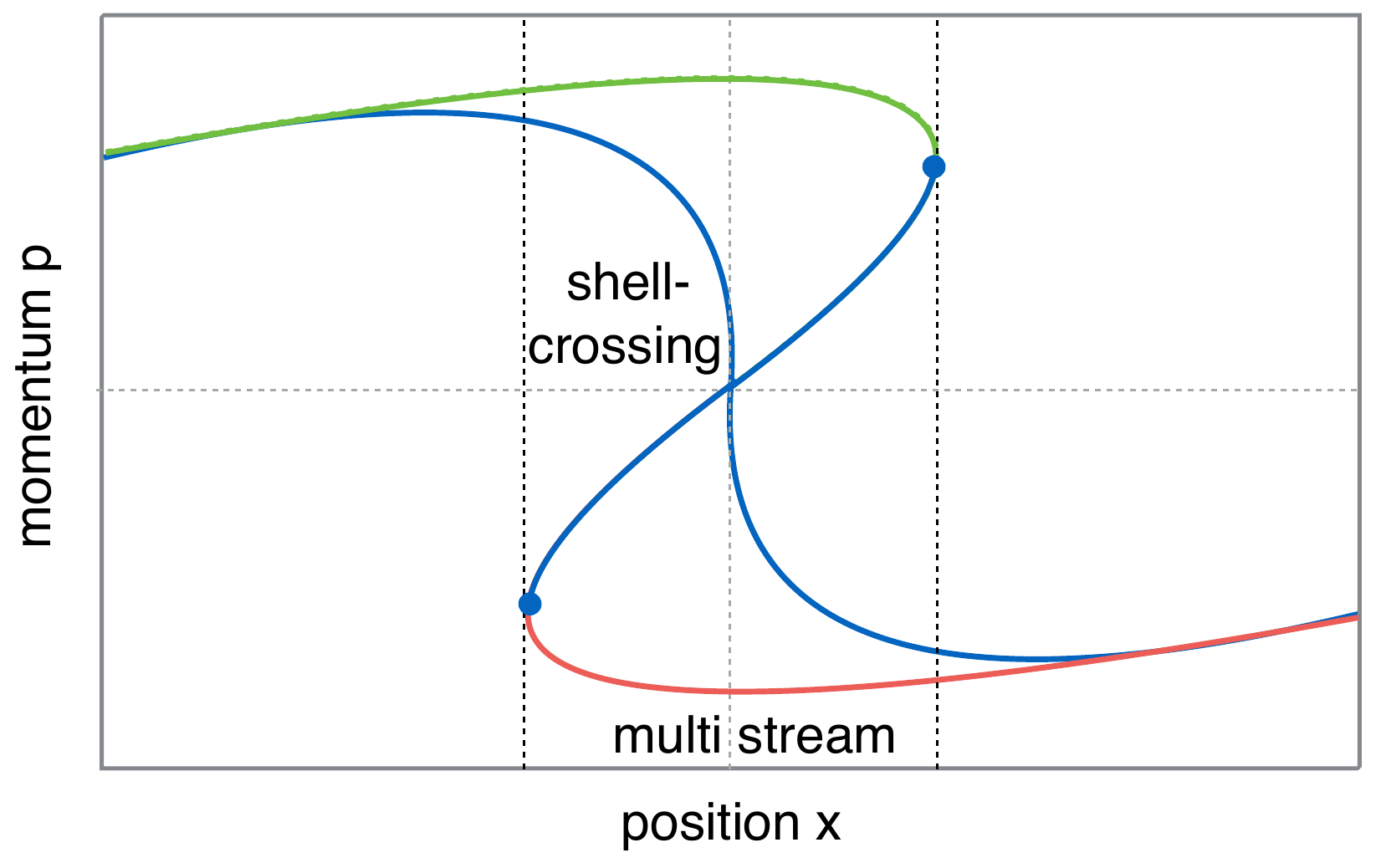} 
\includegraphics[width=0.32\textwidth]{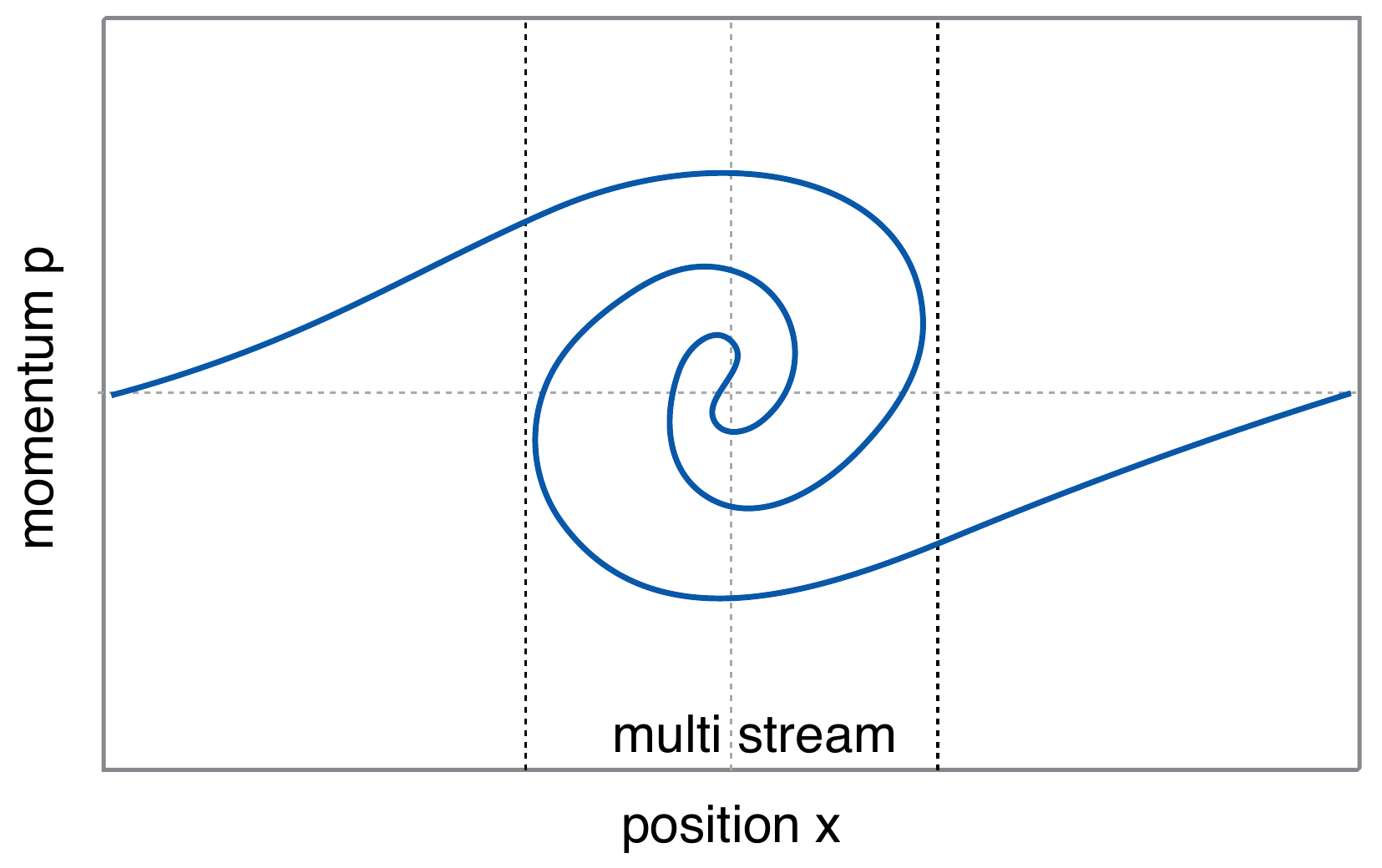} 
\caption{Schematic sketch of the time evolution of cold dark matter in $(1+1)$-dimensional phase-space. 
{\it Left panel:} Due to cold initial conditions the phase-space sheet is initially flat and slowly starts to bend due to the coherent infall caused by gravitational interaction. In this single-stream regime cold dark matter is well-described by a perfect pressureless fluid. 
{\it Middle panel:} During shell-crossing the particle trajectories cross such that the single-stream splits into three fluid streams. 
{\it Right panel:} This process will happen repeatedly and result in a wound up phase-space sheet.}
\label{fig:pheno}
\end{figure}

\paragraph*{Phenomenology of time evolution}

In Figure~\ref{fig:pheno} we show a sketch of the qualitative features of the time evolution for the formation of a bound dark matter structure starting from cold initial conditions. 
A density distribution which is initially almost uniform in a finite region of space, or equivalently a large ensemble of particles uniformly distributed in this region, will create a gravitational potential that leads to a coherent infall to the central region. The particles accelerate and acquire velocities directed towards the center to which they move uniformly. Due to their nonzero velocities at the time they reach the center, the particles overshoot which creates the characteristic S-shape of the phase-space sheet. After passing the center, the particles slow down due to the gravitational potential which pulls them towards the center again. Hence, their velocities decrease and eventually reverse direction to fall into the center again. This corresponds to a rotation of the S-shape around the center where the inner part of the S-shape now resembles the single-streaming stage of early evolution. After a series of shell-crossings, the phase-space sheet will have wound up to an apparent whirl, but without any tears or self-intersections. At some point the phase-space sheet is so tightly wound up, that it becomes effectively stationary and a bound structure has emerged. In general, collisionless self-gravitating systems are expected to evolve towards a steady state after a strong mixing phase such as violent relaxation \cite{Lynden-Bell67}. 

\subsection{The Vlasov hierarchy for moments and cumulants}
\label{sec:Hierarchy}
In practice, one is usually interested in the time evolution of properties of the spatial distribution, especially the density and mass-weighted velocity of the flow, rather than the fully fledged phase-space information encoded in the Vlasov equation. For this purpose, one extracts the relevant information from the phase-space distribution by computing momentum-weighted averages of the phase-space distribution, the so-called moments (with respect to momentum).
\paragraph*{Generating functional for moments and cumulants} The moments $M^{(n)}$ of the phase-space distribution function $f(\vx,\vp)$ are tensorial quantities that can be obtained from the generating functional $G(\vJ)$, which is a linear functional of the phase-space distribution $f$, by taking functional derivatives. In a similar way, the cumulants $C^{(n)}$ resemble the connected parts of the moments and can be determined from the natural logarithm of the generating functional. They provide a good way to understand the prominent perfect pressureless fluid model which is the only consistent truncation of the Vlasov hierarchy, as we discuss in Section~\ref{subsec:fluid}. The generating functional, moments and cumulants are given by 
\begin{subequations}
\begin{align}
&G(\vJ) = \int \vol{3}{p} \exp\left(\tfrac{i}{m}\vp\cdot\v{J}\right) f \,, \label{genfun}\\
&M^{(n)}_{i_1 \cdots i_n}:=\int \vol{3}{p} \frac{p_{i_1}}{m} \ldots \frac{p_{i_n}}{m} f = (-i)^n \left.\frac{\del^n G(\vJ)}{\del J_{i_1} \ldots \del J_{i_n}} \right|_{\v{J}=0} \,, \label{moments}\\
&C^{(n)}_{i_1 \cdots i_n}:= (-i)^n \left.\frac{\del^n \ln G(\vJ)}{\del J_{i_1} \ldots \del J_{i_n}} \right|_{\v{J}=0} \label{cumulants}\,.
\end{align}
\end{subequations}
Note that one useful property of both moments and cumulants is the total symmetry among all their indices.
If the moment/cumulant generating function exists, the probability distribution is uniquely determined by it \cite{LukacsBook}. Note that this does not necessarily mean that the moments/cumulants uniquely determine the probability distribution, because there are cases where all moments exist and yet the limit that defines the generating functions does not exist. The lognormal distribution is such an example.

Note that there are no distributions for which the cumulant generator is a polynomial in $\vJ$ unless one has a Gaussian-like distribution which corresponds to a perfect pressureless fluid and describes the dynamics before shell-crossing. This suggests that for following multi-streaming behaviour that is generated dynamically rather than implemented in the initial conditions (increasing the number of initial streams), the cumulant generator is more informative.

\paragraph*{Vlasov hierarchy} The evolution equation for the moment generator, that simply corresponds to a Fourier transformation of the phase-space distribution with respect to momentum, can be readily obtained from the Vlasov equation and reads
\begin{align}
\del_t G(\vJ,\vx) = \frac{i}{a^2} \vnabla_J\cdot\vnabla_x G - iG \vJ \cdot\vnabla_x V 
\end{align}
Similarly, we get for the cumulant generating function
\begin{align}
\label{eq:evolutionlnG}
\del_t \ln G(\vJ,\vx) = \frac{i}{a^2} \left(\vnabla_J\cdot\vnabla_x \ln G + \vnabla_J \ln G \cdot\vnabla_x \ln G\right) - i\vJ\cdot\vnabla_x V\,.
\end{align}
To obtain evolution equations for moments and cumulants, respectively, one now expands the moment generator $ G(\vJ,\vx)$ or the cumulant generator $\ln G(\vJ,\vx)$ in a power series in $\v{J}$. The evolution equations for the moments $M^{(n)}$ of the phase-space distribution $f$ read
\begin{align}
\label{eq:VlasovHierarchyMoments}
\partial_t M^{(n)}_{i_1 \cdots i_n} &=   - \frac{1}{a^2} \nabla_j M^{(n+1)}_{i_1 \cdots i_n j}  - \nabla_{(i_1} V \cdot M^{(n-1)}_{i_2 \cdots i_n)} \,,
\end{align}
where indices enclosed in round brackets imply symmetrisation according to $a_{(i}b_{j)}=a_ib_j+a_jb_i$. This equation can also be obtained from the Vlasov equation \eqref{VlasovEq} by multiplication with a certain power of momenta $p_{i_1}\cdots p_{i_n}$ and subsequent integration over momentum. It turns out that a coupled Vlasov hierarchy for the moments emerges, which means that in order to determine the time evolution of the $n$-th moment, the $(n+1)$-th moment is required. This closure problem for the hierarchy becomes more transparent when looking at the dynamical equation for the $n$-th cumulant $C^{(n)}$. The time evolution can be determined from the generating functional \eqref{genfun} using the Vlasov equation \eqref{VlasovEq} and reads
\begin{align}
\label{eq:VlasovHierarchyCumulants}
\del_t C^{(n)} _{i_1\cdots i_n} &= -\frac{1}{a^2} \Bigg\{ \nabla_j C^{(n+1)}_{i_1\cdots i_n j} +\sum_{S\in \mathcal P(\{i_1,\cdots,i_n\})} C^{(n+1-|S|)}_{l\notin S,j} \cdot \nabla_j C^{(|S|)}_{k\in S} \Bigg\}  - \delta_{n1} \nabla_{i_1}V \,,
\end{align}
where $S$ runs through the power set $\mathcal P$ of indices $\{i_1,\cdots,i_n\}$ and the Kronecker $\delta_{n1}$ in last term ensures that the potential contributes only to the equation for the first cumulant $C^{(1)}$ describing velocity. 

\paragraph*{Equations for density and velocity}
When applying the cumulant equation~\eqref{eq:VlasovHierarchyCumulants} to the zeroth cumulant describing the logarithm of the density $C^{(0)}=\ln \rho$ and the first cumulant describing the velocity $C^{(1)}_i=v_i$, one finds
\begin{align}
\label{eq:continuity}
\partial_t \ln n &= \frac{-1}{a^2}\left[\vnabla \cdot\vv + \vv\cdot \vnabla \ln n \right] \ \Leftrightarrow \ \partial_t n = \frac{-1}{a^2}\vnabla \cdot (n\vv)\,,\\
\label{eq:Euler}
\partial_t \vv &= \frac{-1}{a^2} \Bigg\{ (\vv\cdot\vnabla)\vv + \underbrace{\vnabla\cdot C^{(2)} + C^{(2)}\cdot \vnabla\ln n}_{\frac{\vnabla\cdot \left[n C^{(2)}\right]}{n}}\Bigg\} - \vnabla V\,,
\end{align}
which are the continuity equation and an Euler-like equation for a general velocity dispersion tensor $C^{(2)}$. Note that our definition of velocity is related to the conjugate momentum and hence not a peculiar velocity. It is often convenient to decompose the velocity into a gradient of a velocity potential and a vector potential. We can derive an evolution equation for the vorticity $\vw=\vnabla\times\vv$. For this it is useful to rewrite $(\vv\cdot\vnabla)\vv = \frac{1}{2}\vnabla{v^2} + \vw\times \vv$
\begin{align}
\label{eq:vorticity}
\partial_t\vw= \frac{-1}{a^2} \Bigg\{ \vnabla\times(\vv\times\vw)+ \vnabla\times \change{\left[\vnabla\cdot \v{C}^{(2)} + \v{C}^{(2)}\cdot\vnabla\ln n\right]}\Bigg\}\,.
\end{align}

\section{Strategies for cumulant hierarchies by the example of Vlasov-Poisson}
\label{sec:FGC}

In principle it would be desirable to adopt an ansatz for the phase-space distribution $f(\vx,\vp)$ that is as general as possible. Due to the complexity of the nonlinear time evolution this is however virtually intractable. Hence, we want to find a simplified model that still captures the phenomenology of the time evolution that leads to halo formation, in particular the formation of multiple streams. Usually, one starts from rather simple initial conditions, corresponding to a perfect pressureless fluid with a given density and an irrotational velocity\change{,} that is described by a gradient field. Using this as an intuition, we will show how one can {\it construct} a model that has the same number of functions but can also capture phenomena of multi-streaming that are induced by the gravitational collapse of the system.

\subsection{Truncation of cumulants: fluid-based models}
\label{subsec:fluid}
A truncation of cumulants at order $m$ is performed by setting all cumulants at order $m$ or higher to zero, $C^{(n\geq m)}\equiv 0$, in the evolution equations for the lower-order cumulants $C^{(n < m)}$ and neglecting possible time evolution of the higher-order cumulants. The truncated hierarchy then consists of closed evolution equations for the lower-order cumulants $C^{(n < m)}$.
One distinguishes between truncations that are {\it consistent} with time evolution, where the evolution equations for higher-order cumulants reduce to $\partial_t C^{(n\geq m)}=0$ and hence vanishing cumulants remain zero, and those which are not.

\subsubsection{Truncation at second order: the perfect pressureless fluid model}

From equation \eqref{eq:VlasovHierarchyCumulants} it appears that one can set $C^{(n\geq 2)} \equiv 0 $ in a consistent manner since each summand in the evolution equation of $C^{(2)}$ contains a factor of $C^{(n\geq 2)}$. This also means that when $C^{(2)}=0$ initially, this property will be preserved until shell-crossing which dynamically produces multiple streams, vorticity and velocity dispersion along with higher cumulants, see \cite{PS09,Hahn14}. The breakdown of the perfect pressureless fluid model description is signalled by the emergence of \change{a} singular density at the instant of shell-crossing. Note that the occurrence of this singularity can be rephrased when going to Lagrangian coordinates, hence using initial positions $\vq$ as labels of the fluid elements and following their displacement $\Psi(\vq,t)$ over time. The Eulerian positions are then obtained by $\vx=\vq + \Psi(\vq,t)$ and the mapping $\vq \rightarrow \vx$ is only one-to-one before shell-crossing\change{,} whereas fluid elements with different initial positions $\vq$ can end up at the same final position $\vx$ in the multi-stream regime such that the mapping from Lagrangian to Eulerian coordinates is not injective anymore. 

The truncation at second order leads to the following moments and cumulants
\begin{subequations}
\begin{align}
C_\d^{(0)} &= \ln \rho_\d\,, \quad \ {C_{\d,i}}^{(1)}=v^{\d}_i  \,, \quad \quad \  {C_\d}^{(n \geq 2)}_{i_1\cdots i_n} = 0\,,\\
M_\d^{(0)} &= \rho_\d\,, \qquad {M_\d}^{(1)}_i=\rho_\d v^{\d}_i\,, \quad { M_\d}^{(n \geq 2)}_{i_1\cdots i_n} = \rho_\d v^{\d}_{i_1} \cdots v^{\d}_{i_n} \,.
\end{align}
\end{subequations}
From the finite number of cumulants, one can easily obtain the cumulant generator 
\begin{align}
\ln G_\d(\vJ)=\ln \rho_\d + i\vJ\cdot\vv_\d \ \Rightarrow \
G_\d(\vJ) =\rho_\d \exp\left[i\v{J}\cdot\vv_{\d}\right] \,,
\end{align}
which is manifestly linear in $\v{J}$, all cumulants of order higher than one vanish identically. This corresponds to a distribution function of a perfect pressureless fluid
\begin{align}
\label{eq:fdust}
f_{\d}(\vx,\vp)=\rho_{\d}(\vx) \delta_D\left(\vp - m\vv_{\d}\right)\,,
\end{align}
which does not include effects like velocity dispersion, encoded in the second cumulant $C^{(2)}$ or higher-order effects. 
Therefore for the dust ansatz $f_\d$, the Vlasov equation is equivalent to its first two equations of the hierarchy of cumulants, the pressureless fluid system consisting of the continuity equation~\eqref{eq:continuity} and Euler equation~\eqref{eq:Euler}. Since the perfect pressureless fluid has zero velocity dispersion, $C^{(2)}\equiv 0$, an initially irrotational velocity will stay irrotational until shell-crossing. Hence, one can set $\vv=\vnabla\phi_\d$ and reduce the Euler equation \eqref{eq:Euler} for the velocity to the Bernoulli equation for the velocity potential $\phi_\d$ such that the perfect pressureless fluid system reads
\begin{subequations}
\label{Fluid}
\begin{align}
\del_t \rho_\d &=   -\frac{1}{a^2}\vnabla\cdot( \rho_\d \vnabla\phi_\d) \,, \label{eq:continuityfl}\\
\del_t \phi_{\d} &= -\frac{1}{2a^2} (\vnabla\phi_\d)^2- V   \,. \label{eq:Bernoullifl}
\end{align}
\end{subequations}
If $\rho_\d$ and $\vv_\d$ fulfill these equations and higher cumulants are set to zero identically, then all evolution equations of the higher moments are automatically satisfied. 

\subsubsection{Problems with higher-order truncations and multiple fluids}
Since, according to \eqref{eq:VlasovHierarchyCumulants}, the time evolution of $C^{(n\geq 3)}$ is sourced also by summands containing solely $C^{(2)}$, it cannot be trivially fulfilled when setting $C^{(n\geq 3)} \equiv 0$. A similar reasoning applies to all higher cumulants $C^{(n\geq 3)}$ and demonstrates that there is no consistent truncation of the hierarchy of cumulants apart from the one at second order. These arguments are seconded by numerical evidence indicating that as soon as velocity dispersion encoded in $C^{(2)}$ becomes relevant, even higher cumulants are sourced dynamically, see \cite{PS09}. Therefore studying truncations at higher-order does not seem promising for understanding gravitationally induced multi-streaming phenomena that lead to halo formation. For a perturbative treatment of the backreaction effect of stream-crossing on large-scales in terms of an effective field theory on large scales see however \cite{Aviles16,McDonald18}.

In fact, the perfect pressureless fluid model is the only consistent cumulant truncation of the Vlasov hierarchy, for that special case the generating functional $\ln G(\vJ)$ is a polynomial in $\vJ$. This can be connected to a much more general result of \cite{Marcinkiewicz39} stating that the normal distribution is the only probability distribution whose cumulant generating function is a polynomial, i.e. having a finite number of non-zero cumulants. 

An alternative to truncating the series of cumulants is to reduce the complexity of phase-space models by using \change{fluid-based} approximations. 
In the method of moments \cite{McGraw97,Yuan11}, higher moments are represented as functions of lower moments and certain quadratures are used to obtain estimates of the phase-space distribution. 
An intuition of the underlying principle can be obtained by considering a multi-fluid phase-space distribution $f(\vx,\vp)=\sum_{i=1}^N f_\d^i(\vx,\vp)$ with $N$ fluids $f_\d$ of the form  \eqref{eq:fdust} with different densities $\rho_d^i$ and single-stream velocities $\vv_\d^i$. This induces an infinite number of cumulants, but since there is a finite number of free functions $\{\rho_d^i(\vx),\vv_\d^i(\vx)\}_{i=1,\ldots,N}$, higher moments (and hence cumulants) are represented as functions of lower moments in a recursive fashion. For the purpose of starting from a single perfect fluid and capturing a series of shell-crossings that causes a successively increasing number of streams, this method does not seem promising as it remains unclear how new streams are created dynamically.

\subsection{Finitely generated cumulants}
In the last paragraph, we have seen that a finite number of cumulants cannot be an adequate description of the phase-space probability distribution that describes the nonlinear gravitational evolution of dark matter that leads to the formation of bound structures. We can however generalise the idea of a simplified description of the whole phase-space distribution in terms of cumulants that are functions in position space. 

\subsubsection{Probability distributions of one variable} 
For simplicity, let us first consider a probability distribution function $\mathcal P(y)$ which is just a function of one scalar variable $y$ (rather than two vectors as in our case). It has a moment/cumulant generating function that is a scalar function $G(J)=\int \exp(iJy) \mathcal P(y)\, dy$ and can be expanded in terms of the cumulants $C_n=\partial^n \ln G/\partial (iJ)^n$ (note that here $C_0=1$ due to normalisation) which are just real numbers (rather than tensorial functions of space). Unless the distribution is  Gaussian, this Taylor-like series is infinite. However, it could still be simplified by cumulants that have a fixed functional form with a finite number of parameters or a cumulant recursion relation. In order to go from truncation to finitely generated cumulants, one replaces the idea of a `finite number of cumulants' to a `finite number of generators for cumulants' \cite{PistoneWynn99}. Let us make some examples:
\begin{itemize}
\item The continuous exponential distribution $P(y|\lambda)=\lambda\exp(-\lambda y) \Theta(y)$, with the Heaviside step function $\Theta$, has cumulants $C_n=\lambda^{-n} (n-1)!$ that are all specified by just one parameter $\lambda$.
\item The probability $\mathcal P(\rho|n,\sigma(R,t))$ of finding a nonlinear dark matter density $\rho$ in a sphere of sufficiently large radius $R$ at  cosmic time $t$ that evolved from Gaussian initial conditions is almost entirely determined by two parameters. One parameter for the scale-dependence of the initial conditions, namely the slope $n$ of the initial power spectrum and another parameter for the variance of the nonlinear density $\sigma(R,z)$. The reason behind this is that there is an approximate nonlinear mapping between linear and nonlinear densities in spheres and their radii that stems from large deviation statistics \cite{Bernardeau94,Uhlemann16}. In practice, the resulting transformations are close to logarithmic or of power-law (Box-Cox) type depending on the value of $n$ \cite{Szapudi13}. 
\item The discrete Poisson distribution, $\mathcal P(N|\lambda)=\lambda^N \exp(-\lambda)/N!$ has cumulants that are all identical and equal to the expectation value $C_n=\lambda$. It can for example be used to relate continuous densities in spheres $\rho$ to discrete counts of objects $N$ such as galaxies \cite{Szapudi2004}.
\end{itemize}

\subsubsection{Phase-space probability distributions} 
Let us boldly generalise the idea of finitely generated cumulants to a probability distribution of two vector variables. This means that all cumulants are now tensorial functions of position (rather than numbers) and should be functionals (rather than functions) of a finite number of base functions (rather than base parameters). In particular, one can anticipate that one needs functionals that can generate a higher-order tensor structure from lower-order ones, such as derivative operators. 

Since initially only the lowest two cumulants, $\{C^{(0)},C^{(1)}_i\}$, are present, those two cumulants should be the base functions from which higher-order cumulants are dynamically generated. Hence, one could hope to find a recurrence relation for cumulants that allows to describe the higher-order cumulants as functionals of lower-order cumulants. This idea can be used to construct a viable ansatz for the phase-space distribution $f(\vx,\vp)$ as a generally nonlocal functional of the lower-order cumulants $f_T[C^{(0)},C_i^{(1)}](\vp)$, hence reducing dynamics from 6-dimensional phase-space down to 3-dimensional position space. Since our formulation is based on cumulants, we will consider the cumulant generator $\ln G_T[C^{(0)},C_i^{(1)}](\vJ)$ instead of the phase-space distribution $f_T$ itself.  
The key requirement for a phenomenologically viable ansatz for the phase-space distribution is that the functional form is preserved through time evolution and hence captures the effects of nonlinear gravitational dynamics.  

We put our emphasis here on the phase-space dynamics with given initial conditions, having in particular halo formation in mind that requires a dynamical framework that can treat multi-streaming. If one, in contrast, is only interested in the statistical properties of the phase-space distribution, averaged over (typically Gaussian) initial conditions rather than the precise dynamics for specific initial conditions, one might benefit from field theory methods described in \cite{Valageas04}.

\section{Deriving Schr\"odinger-Poisson using finitely generated cumulants}
\label{sec:SPder}
As informative and instructive example for the usage of finitely generated cumulants for deriving approximate closed-form system for cumulant dynamics, we derive the Schr\"odinger method for the Vlasov-Poisson equation. The recipe of our procedure can be summarised in three steps
\begin{enumerate}
\item Choose lower-order cumulants as base functions and restrict the functional form of the cumulant generator, guided by the initial conditions and the short-term evolution.
\item Determine the functional form of the velocity dispersion tensor entering the evolution of lower-order cumulants that is compatible with the assumptions.
\item Rewrite the time evolution equation of the cumulant generator in terms of the time evolution of base-functions to infer the form of the functionals.
\end{enumerate}

\subsection{Base assumptions for cumulant generator} 
Let us assume that the base functions for building all cumulants are the density $n(\vx)$ and velocity potential $\phi(\vx)$. They determine the lowest two cumulants via $C^{(0)}=\ln n$ and $C^{(1)}_i=\nabla_i \phi$ and hence the Taylor-expansion of our trial cumulant generator at first order
\begin{align}
\ln G_T(\vJ)= \ln n + i\vJ\cdot \vnabla \phi + \mathcal O(\vJ^2)\,.
\end{align}
Now, using the idea of finitely generated cumulants, all cumulants and hence the whole cumulant generator should be built as a functional of those two base functions $\ln G(\vJ)[n,\phi]$. While in principle this functional could be very complicated, let us assume that it is \textit{linear} in the base-functions\footnote{Note that the assumption of linearity is mainly motivated by simplicity. Our aim is to find a suitable ansatz that approximately solves the cumulant evolution equation~\eqref{eq:evolutionlnG} by virtue of evolution equations for the lowest order moments. If the trial ansatz for the cumulant generator is nonlinear in the basis functions, partial derivatives in \eqref{eq:evolutionlnG} need to be evaluated with chain and product rules, thus generating a cascade of additional nonlinear terms. This would make it very challenging to infer the properties of the functionals.} and \textit{only depends on $\vJ$}.  This means it can be decomposed into two linear operators, $\mathcal  O_n$ and $\mathcal O_\phi$, acting on the lowest order cumulants, respectively
\begin{align}
\label{eq:lnGoperatornphi}
\ln G_T(\vJ)[n,\phi] &= \mathcal O_n(\vJ) \ln n + i\mathcal O_\phi(\vJ) \phi \,.
\end{align}

Since the first two cumulants shall be $\ln n$ and $\vnabla \phi$, the first two Taylor coefficients of the operators are fixed to be
\begin{align}
\label{eq:lnGoperatornphiinicond}
\mO_n(\vJ=0)&=\mathbb{1} \,,\quad \vnabla_J \mO_n(\vJ=0)=0\,,\\
\mO_\phi(\vJ=0)&= 0 \,,\quad \vnabla_J \mO_\phi(\vJ=0)=\vnabla_x\,.
\end{align}

Our goal is to determine the linear operators, $\mO_n$ and $\mO_\phi$, such that the evolution equations for the cumulant generator, and hence the higher cumulants induced by this operator are approximately automatically fulfilled given the evolution equation for $\ln n$ and $\phi$.

\subsection{Time evolution for base functions} 
Since our base functions are the log-density $\ln n$ and the velocity potential $\phi$, we need evolution equations for those fundamental degrees of freedom. The first one is the continuity equation~\eqref{eq:continuity} written for a gradient field
\begin{align}
\label{eq:continuityphi}
\partial_t \ln n &= \frac{-1}{a^2}\left[\Delta \phi + \vnabla \ln n \cdot \vnabla \phi\right] \ \Leftrightarrow \ \partial_t n = \frac{-1}{a^2}\vnabla \cdot (n\vnabla\phi)\,,
\end{align}
while the second one should be an analogue of the Bernoulli equation~\eqref{eq:Bernoullifl},
\begin{align}
\label{eq:Bernoulli}
\del_t \phi &= -\frac{1}{a^2} \left\{\frac{1}{2}\left(\vnabla\phi\right)^2  + \tilde C^{(2)}\right\} -V\,,
\end{align}
including the effects of nonzero velocity dispersion in the scalar term $\tilde C^{(2)}$.

First, we need to check whether for an initially potential flow $\vv=\vnabla\phi$ the Bernoulli equation~\eqref{eq:Bernoulli} is equivalent to the Euler equation~\eqref{eq:Euler}. Given the time evolution equation~\eqref{eq:vorticity} for vorticity, we observe that an initially potential flow only stays potential if the term in square brackets is zero, which means
\begin{align}
\label{eq:potentialflow}
\epsilon_{lki} \left[ \nabla_k \nabla_j\cdot C^{(2)}_{ij} + \nabla_k C^{(2)}_{ij}\nabla_j\ln n + C^{(2)}_{ij}\nabla_k\nabla_j\ln n\right]\equiv 0\,.
\end{align}
Since $\epsilon$ is totally antisymmetric, this can be guaranteed if the term in brackets is symmetric in $k\leftrightarrow i$, or identically zero which corresponds to the perfect pressureless fluid with vanishing higher cumulants $C^{(n\geq 2)}\equiv 0$. The first term cannot be cancelled by the other two nonlinear terms, since we assumed that the cumulants are linear in the log-density and velocity potential. Hence, we need the first term to be symmetric in $k\leftrightarrow i$ and due to the symmetry of the indices $C^{(2)}_{ij}=C^{(2)}_{ji}$, we can infer $C^{(2)}_{ij}(\vx)\propto \nabla_i \nabla_j g(\vx)$ with some function $g(\vx)$. This form also renders the second term symmetric in $k\leftrightarrow i$, while from the last term one can deduce
\begin{align}
\label{eq:C2lnn}
C^{(2)}_{ij}(\vx)&=c \nabla_i \nabla_j \ln n(\vx) \,,
\end{align}
with some real constant $c$.
For this case, one can rewrite the $C^{(2)}$ terms in the Euler equation~\eqref{eq:Euler} as a total derivative of the scalar quantity $\tilde C^{(2)}$ with prefactor $c$%
\begin{align}
\tilde C^{(2)} &: = c\left(\Delta \ln n + \frac{1}{2}(\vnabla \ln n)^2\right) = 2c\frac{\Delta \sqrt{n}}{\sqrt{n}}\,,\\
\label{eq:C2andC2tilde}
\nabla_i\tilde C^{(2)}&=\nabla_jC^{(2)}_{ij} + C^{(2)}_{ij}\nabla_j\ln n =\frac{\nabla_j[nC^{(2)}_{ij}]}{n}\,.
\end{align} 
In order to ensure a positive velocity dispersion $\sigma^2=C^{(2)}_{ii}$ on large scales one has to choose a negative $c$ because $\ln n$ is typically concave, i.e. has negative curvature because the density $n$ is peaked around the points of shell-crossing\change{,} which create velocity dispersion. Now our Bernoulli equation becomes
\begin{align}
\label{eq:Bernoullipot}
\del_t \phi &= \frac{-1}{a^2} \left\{\frac{1}{2}\left(\vnabla\phi\right)^2  +  c\left(\Delta \ln n + \frac{1}{2}(\vnabla \ln n)^2\right) \right\} -V \,.
\end{align}
Note that, while our evolution equation is for a scalar $\phi$ field, this does not mean that the velocity field stays a gradient at all times. Indeed, the time evolution can induce jumps in $\phi$ which coincide with the occurrence of multi-streaming and generate vorticity through circulation around the domains of shell-crossing \cite{Kopp17}, where Kelvin's circulation theorem is violated \cite{Hankel1861,Kelvin1869,Villone17}
\begin{align}
\varointctrclockwise_C \vnabla\phi \cdot d\vx \neq 0\,.
\end{align}

At this stage, one can already notice that from a special ansatz of finitely generated cumulants, one obtains a Bernoulli equation \eqref{eq:Bernoullipot} with an additional dispersion term, compared to the perfect fluid form. The role of this extra term as a special form of velocity dispersion becomes more apparent when writing down the equation for the velocity $\vv=\vnabla\phi$ obtained from \eqref{eq:Bernoullipot}, which gives the Euler-like equation \eqref{eq:Euler} with the special velocity dispersion \eqref{eq:C2lnn}. This term is called quantum velocity dispersion (or quantum-pressure) because it is obtained when the Schr\"odinger-Poisson equation is rewritten in its fluid form using the Madelung transformation \cite{Madelung27} and setting $c=-\hbar^2/(4m^2)$. Indeed, we can reverse this procedure by introducing the function $\psi=\sqrt{n}\exp(i\phi/\hbar)=\exp\left(\frac{1}{2} \ln n + \frac{i}{\hbar} \phi\right)$, whose time evolution can be shown to follow the Schr\"odinger-Poisson equation
\begin{align}
\partial_t \psi = \left(\frac{1}{2}\partial_t \ln n + \frac{i}{\hbar}\partial_t \phi\right)\psi \quad \stackrel{\eqref{eq:continuityphi},\eqref{eq:Bernoullipot}}{\Longrightarrow} \quad i\hbar\partial_t \psi =\frac{-\hbar^2}{2a^2m^2}\Delta\psi + V\psi\,,
\end{align}
where we have used the continuity equation~\eqref{eq:continuityphi} for $\partial_t \ln n$ and the special Bernoulli equation~\eqref{eq:Bernoullipot} for $\partial_t \phi$.

So far, we have determined the form of the velocity dispersion which is the first of the higher-order cumulants that arise from the multi-streaming induced by shell-crossing. We have seen that we obtain a closed fluid-like system of continuity and Bernoulli equation from our finitely generated cumulant ansatz, which can be rewritten in terms of a Schr\"odinger-Poisson equation. Next, in order to obtain an expression for a phase-space distribution that is close to Vlasov-Poisson, we will  look at the full hierarchy of cumulants.

\subsection{Time evolution for cumulant generator} 
Since the operators $\mO_n$ and $\mO_\phi$ were assumed to solely depend on $\vJ$, time and spatial derivatives only act on the base functions
\begin{align}
\partial_t \ln G_T &= \mO_n\partial_t \ln n + i\mO_\phi\partial_t \phi\,,\quad
\vnabla_x \ln G_T =\mO_n \vnabla_x \ln n + i\mO_\phi \vnabla_x \phi\,.
\end{align}
To check whether the desired Vlasov-like evolution equation for the cumulant generator $\ln G$ can be approximately fulfilled by our finitely generated cumulant ansatz for $\ln G_T$, we plug \eqref{eq:lnGoperatornphi} into \eqref{eq:evolutionlnG} to obtain
\begin{align}
 &\mathcal O_n \partial_t \ln n + i\mathcal O_\phi \partial_t \phi + i\vJ\cdot\vnabla V\\
 \notag &= \frac{i}{a^2}\Bigg\{\vnabla_J (\mathcal O_n) \vnabla_x \ln n + i\vnabla_J (\mathcal O_\phi) \vnabla_x \phi + \vnabla_J\left( \mathcal O_n \ln n + i\mathcal O_\phi \phi\right)\left[ \mathcal O_n\vnabla_x \ln n + i\mathcal O_\phi \vnabla_x \phi\right] \Bigg\}\,.
\end{align}

Now we can use the continuity equation~\eqref{eq:continuityphi} for $\partial_t \ln n$ and the Bernoulli-like equation~\eqref{eq:Bernoullipot} for $\partial_t \phi$ to obtain 
\begin{align}
\label{eq:consistencyeq}
 &-\mathcal O_n \left[\vnabla_x^2 \phi + \vnabla_x \ln n \cdot \vnabla_x \phi\right] - i\mathcal O_\phi\left[\frac{1}{2}(\vnabla_x\phi)^2+c\left(\vnabla_x^2 \ln n + \frac{1}{2}(\vnabla_x \ln n)^2\right) + a^2V\right]\!\!\\
\notag &= i\Bigg\{\vnabla_J \mathcal O_n \cdot\vnabla_x \ln n + i\vnabla_J \mathcal O_\phi \cdot\vnabla_x \phi + \vnabla_J \left( \mathcal O_n \ln n + i\mathcal O_\phi \phi\right)\cdot \left[ \mathcal O_n\vnabla_x \ln n + i\mathcal O_\phi \vnabla_x \phi\right] \\
\notag &\qquad- a^2 \vJ\cdot\vnabla_x V\Bigg\}\,.
\end{align}
If we focus on the terms which are linear in $\phi$ and $\ln n$, respectively, we can read off properties that relate the two operators $\mO_n$ and $\mO_\phi$
\begin{align}
\label{eq:propertiesOnOphi}
\mO_n \vnabla_x^2 \phi&=  \vnabla_J (\mO_\phi)\cdot \vnabla_x \phi\ \Rightarrow \ \vnabla_J (\mathcal O_\phi)= \mathcal O_n \vnabla_x\,,\\
 -c\mO_\phi\vnabla_x^2\ln n&= \vnabla_J(\mO_n)\cdot \vnabla_x\ln n\Rightarrow \ \vnabla_J (\mathcal O_n) = -c \mO_\phi \vnabla_x\,.
\end{align}
Hence, we see that $\vJ$-derivatives on the operators must correspond to spatial derivatives on the base functions 
\begin{align}
\vnabla_J \vnabla _J \mathcal O_{n/\phi}=-c \mathcal O_{n/\phi} \vnabla_x\vnabla_x\,,
\end{align}
with `boundary' conditions for $\vJ=0$ \eqref{eq:lnGoperatornphiinicond} and \eqref{eq:C2lnn}. From the previous step we also know that we need $c<0$ to ensure a physically meaniningful positive velocity dispersion. Hence, we choose $c=-\tilde c^2$ and find the solution
\begin{align}
\label{eq:solOnOphi}
(\mO_n,\mO_\phi)(\vJ)=\left(\cosh(\tilde c\vJ\cdot\vnabla_x) ,\frac{1}{\tilde c}\sinh(\tilde c\vJ\cdot\vnabla_x)\right) \,.
\end{align}

Now we need to check whether our trial solution \eqref{eq:solOnOphi}, obtained from the linear terms, also equates the terms which are nonlinear in the base functions.
From the quadratic terms in $\phi$ and $\ln n$ one obtains
\begin{align}
\mO_\phi (f^2) \stackrel{!}{=} 2 (\mO_n f) (\mO_\phi f) \text{ with } f\in \{\vnabla\phi,\vnabla\ln n\} \,,
\end{align}
after simplification using the properties \eqref{eq:propertiesOnOphi}, while the mixed term between $\phi$ and $\ln n$ in \eqref{eq:consistencyeq} gives
\begin{align}
\mO_n \left(\vnabla_x\ln n\cdot\vnabla_x\phi\right)&\stackrel{!}{=}\vnabla_J\mO_n \ln n \cdot \mO_\phi\vnabla_x \phi + \vnabla_J\mO_\phi \phi \cdot\mO_n \vnabla_x\ln n\\
&\stackrel{!}{=} -c\mO_\phi\vnabla_x \ln n \cdot \mO_\phi\vnabla_x \phi + \mO_n \vnabla_x\ln n\cdot \mO_n\vnabla_x \phi \,.
\end{align}
One can now verify that both relations are fulfilled using the following identities for exponentiated derivatives (that can be derived using the product rule, series expansions and Cauchy product),
\begin{align}
\label{eq:productruleexp}
\exp  \left(\tilde{\vJ}\cdot \vnabla\right) \left[g(x)h(x)\right] &= \left[ \exp  \left(\tilde{\vJ}\cdot \vnabla\right) g(x)\right] \left[\exp  \left(\tilde{\vJ}\cdot \vnabla\right) h(x)\right]\,,\\
\label{eq:productrulesinh}
\sinh \left(\tilde{\vJ}\cdot \vnabla\right) \left[g(x)^2\right] &= 2\sinh \left(\tilde{\vJ}\cdot \vnabla\right) g(x)\cosh \left(\tilde{\vJ}\cdot \vnabla\right) g(x)\,,\\
\label{eq:productrulecosh}
\cosh \left(\tilde{\vJ}\cdot \vnabla\right) \left[g(x)h(x)\right] &=\left[\sinh \left(\tilde{\vJ}\cdot \vnabla\right) g(x)\right]\left[\sinh \left(\tilde{\vJ}\cdot \vnabla\right) h(x)\right] \\
\notag &\quad + \left[\cosh \left(\tilde{\vJ}\cdot \vnabla\right) g(x) \right]\left[\cosh \left(\tilde{\vJ}\cdot \vnabla\right) h(x)\right] \,.
\end{align}

From the term that includes the gravitational potential, we can deduce that we need
\begin{align}
\label{eq:Schroedingercorr}
(\mO_\phi- \vJ\cdot\vnabla_x) V \simeq 0\,,
\end{align}
which shows that the price for finding a self-consistent closure is that the evolution only approximately corresponds to the Vlasov hierarchy. Note that interestingly, the lowest order cumulant whose evolution equation is affected by the approximate nature of the closure is not the velocity dispersion $C^{(2)}$ (which corresponds to the second order Taylor expansion being quadratic in $\vJ$), but only the third order cumulant $C^{(3)}$. More precisely, the time evolution equation for our finitely generated cumulant ansatz~\eqref{eq:lnGoperatornphi} reads
\begin{align}
\label{eq:evolutionlnGnphi}
\del_t \ln G_T = \frac{i}{a^2} \left(\vnabla_J\cdot\vnabla_x \ln G_T + \vnabla_J \ln G_T \cdot\vnabla_x \ln G_T\right) - \frac{i}{\tilde c} \sinh\left(\tilde c\vJ\cdot\vnabla_x\right) V\,.
\end{align}
The time evolution of this equation is close to the Vlasov evolution if the nonlinear terms in the expansion of the $\sinh$ are suppressed compared to the other terms. While the cumulant ansatz requires a nonzero constant $\tilde c$ to avoid the too simplistic perfect fluid limit, the evolution requires the parameter $\tilde c$ in the solution to the consistency equations \eqref{eq:solOnOphi} to be small in order to correspond to Vlasov-Poisson dynamics. As we will discuss later, this means that one has to consider a semi-classical limit, where $\tilde c$ is kept nonzero but parametrically small compared to the scales on which one seeks to resolve the phase-space distribution. Furthermore, the spatial derivatives of the potential have to be controlled, meaning that the potential $V$ should not exhibit oscillations on small scales of order $\tilde c$. For the gravitational potential determined from the density via the Poisson equation this should be fulfilled given that the inverse Laplacian essentially acts as a smoothing of the density perturbation that sources it. Even if the (approximate) density $n$ devolops order unity oscillations on small scales related to the size of $\tilde c$, the approximate potential converges to the classical one as $\tilde c$ which corresponds to the quantum-scale $\hbar$ or equivalently the inverse mass-scale $m^{-1}$ \cite{Mocz18}. 

The mathematically proper way to quantify the similarity of the time evolution of the underlying systems is to introduce a coarse-graining in phase-space $(\sigma_x,\sigma_p)$ ensuring that scales comparable with the size of the introduced parameter $\tilde c$ are not resolved. We shortly discuss this point in Section~\ref{subseq:VPfromSP}, where we make contact with the common approach based on the quantum-classical correspondence.

\subsection{Result for cumulant generator}
To summarise, we deducted the following cumulant generator based on the idea of finitely generated cumulants~\eqref{eq:lnGoperatornphi}
\begin{align}
\label{eq:cgfcoshsinh}
\ln G_T(\vJ)  &
= \cosh\left(\tilde c\vJ\cdot \vnabla\right)\ln n(\vx)+\frac{i}{\tilde c}\sinh\left(\tilde c\vJ\cdot \vnabla\right)\phi(\vx)\,,
\end{align}
which is a \textit{linear} functional of the two underlying dynamical entities, the log-density $\ln n$ and velocity potential $\phi$, corresponding to the two lowest order cumulants. This linear functional generates cumulants at all orders, with a recurrence relation that relates cumulants of even/odd order to a second derivative of their lower-order counterparts. The explicit expressions can be straightforwardly obtained from \eqref{cumulants} and read
\begin{align}
\label{eq:cumrecrel}
C^{(n+2)}_{i_1 \cdots i_{n+2}} &= -\tilde c^2(C^{(n)}_{i_1\cdots i_{n}})_{,i_{n+1} i_{n+2}} \ \ \forall n\in \mathbb N_0\,,\\ 
C^{(2n)}_{i_1 \cdots i_{2n}}&= \left(-\tilde c^2\right)^{n} (\ln n)_{,i_1 \cdots i_{2n}}= \left(-\tilde c^2\right)^{n} (C^{(0)})_{,i_1 \cdots i_{2n}}\,,\\
C^{(2n+1)}_{i_1 \cdots i_{2n+1}}&= \left(-\tilde c^2\right)^{n}\phi_{,i_1 \cdots i_{2n+1}}=  \left(-\tilde c^2\right)^{n}(C^{(1)}_{i_{2n+1}})_{,i_1 \cdots i_{2n}} \,.
\end{align}
Requiring a small parameter $\tilde c$, we explicitly demonstrated the consistency of the cumulant dynamics {\it at all orders} in a clear and concise way. In Section~\ref{subseq:VPfromSP} it is shown that this is precisely the Schr\"odinger method conjectured before as trick to solve Vlasov-Poisson using quantum-classical correspondence for small values of $\tilde c$ corresponding to $\hbar$ \cite{WK93}. Our result extends upon previous work \cite{Uhlemann14}, where a demonstration of the consistency of the time evolution for the velocity dispersion $C^{(2)}_{ij}$ was presented.

\subsection{Generalised result for cumulant generator with time-dependent parameter}

Let us slightly generalise the result we have just obtained and include a time-dependence in the parameter $\tilde c(t)$ in the ansatz for the cumulant generating function~\eqref{eq:cgfcoshsinh}
\begin{align}
\label{eq:cgfcoshsinhtime}
\ln G_T(\vJ)  = \cosh\left(\tilde c(t)\vJ\cdot \vnabla\right)\ln n(\vx)+\frac{i}{\tilde c(t)}\sinh\left(\tilde c(t)\vJ\cdot \vnabla\right)\phi(\vx)\,.
\end{align}
This induces an extra term of the form $\partial_{\tilde c} \ln G_T \partial \tilde c/\partial t$ into~\eqref{eq:evolutionlnGnphi} with
\begin{align}
\partial_{\tilde c} \ln G_T &= \sinh\left(\tilde c\vJ\cdot \vnabla\right) \vJ\cdot \vnabla \ln n+\frac{i}{\tilde c} \left[\vJ\cdot \vnabla \cosh\left(\tilde c\vJ\cdot \vnabla\right)-\frac{1}{\tilde c}\sinh\left(\tilde c\vJ\cdot \vnabla\right)\right]\phi\\
&= \sum_{n\in 2\mathbb N_0} \frac{{\tilde c}^{n+1}}{(n+1)!}  (\vJ\cdot\vnabla_x)^{n+2} \ln n -i \sum_{n\in 2\mathbb N_0} {\tilde c}^{n+1} \frac{n+2}{(n+3)!} (\vJ\cdot\vnabla_x)^{n+3} \phi(\vx)\,.
\end{align}
This shows that corrections due to a time-dependent $\tilde c$ only enter the time evolution equations for cumulants of order 2 and higher while leaving the closed equations for the base functions $n$ and $\phi$ unchanged.

\subsection{Common approach: Vlasov-Poisson from Schr\"odinger-Poisson}
\label{subseq:VPfromSP}
\subsubsection{Guessing Schr\"odinger-Poisson using quantum-classical correspondence}

In cosmology, the  Schr\"odinger method was first put forward in \cite{WK93,DW96,W97} in the context of numerical simulations of dark matter dynamics, with the aim to replace or complement N-body simulations. The idea originates from the correspondence between phase-space distributions in quantum mechanics and classical mechanics. Essentially, the Schr\"odinger method approximately solves the Vlasov-Poisson equation by making an educated guess for the functional form of the phase-space distribution $f$ using the Wigner function from the phase-space formulation of quantum mechanics
\begin{equation}
\label{eq:Wignerf}
f_W(\vx,\vp) = \int \frac{\,\vol{3}{\tilde x}}{(2\pi)^3}  \exp\left[i\vp\cdot \tilde\vx\right] \bar\psi\left(\vx+\tfrac{\hbar}{2}\tilde \vx\right)\psi\left(\vx-\tfrac{\hbar}{2}\tilde \vx\right)\,,
\end{equation}
with a complex wave-function $\psi$ ($\bar\psi$ denoting its complex conjugate) that evolves on position space. This ansatz shares some simplicity with the perfect pressureless fluid model because the phase-space distribution is entirely described in terms of two functions on position space given as amplitude and phase of the wave-function. The time evolution of the (pseudo-) wave-function is obtained from solving a Schr\"odinger-Poisson equation
\begin{align} 
\label{schrPoissEqFRW}
i\hbar \del_t \psi &= - \frac{\hbar^2}{2a^2m} \Delta \psi + m V\psi  \,,\quad 
\Delta V=\frac{4\pi G\,\rho_0}{a}\Big(|\psi|^2 -  1 \Big) \,.
\end{align}
When splitting the wavefunction in amplitude and phase $\psi = \sqrt n \exp\left(i\phi/\hbar\right)$ one readily obtains the continuity equation~\eqref{eq:continuityphi} and the Bernoulli equation~\eqref{eq:Bernoullipot} containing an extra `quantum' velocity dispersion \cite{Madelung27}.

The Schr\"odinger-Poisson evolution ensures that the Wigner function obeys an equation which is similar to the Vlasov-Poisson equation 
\begin{align}
\label{eq:WignerVlasov}
\partial_t f_\W &= \left[ \frac{\vp^2}{2a^2m}+m V\right]   \frac{2}{\hbar} \sin\left(\frac{\hbar}{2}( \overleftarrow{\vnabla}_{\! \!  x} \cdot\overrightarrow{\vnabla}_{\! \!  p}- \overleftarrow{\vnabla}_{\! \!  p} \cdot \overrightarrow{\vnabla}_{\! \!  x})\right) f_\W \,,
\end{align}
where the Poisson-bracket from the Vlasov-equation~\eqref{VlasovEq} has been replaced by a Moyal-bracket \cite{Moyal49,Groenewold46} that adds higher-derivative terms controlled by the small parameter $\hbar$. 
Note that higher powers of $\overleftarrow{\vnabla}_{\! \!  p} \cdot \overrightarrow{\vnabla}_{\! \!  x}$ do not contribute because the Hamiltonian is only quadratic in momentum and hence higher derivatives vanish identically. This means, we can rewrite 
\begin{align*}
\frac{2}{\hbar}\sin\left(\frac{\hbar}{2}( \overleftarrow{\vnabla}_{\! \!  x} \cdot\overrightarrow{\vnabla}_{\! \!  p}- \overleftarrow{\vnabla}_{\! \!  p} \cdot \overrightarrow{\vnabla}_{\! \!  x})\right) = - \overleftarrow{\vnabla}_{\! \!  p} \cdot \overrightarrow{\vnabla}_{\! \!  x} + \frac{2}{\hbar}\sin\left(\frac{\hbar}{2} \overleftarrow{\vnabla}_{\! \!  x} \cdot\overrightarrow{\vnabla}_{\! \!  p}\right)\,,
\end{align*}
which allows us to recast the Wigner-Vlasov equation in the following form
\begin{align}
\label{eq:WignerVlasov2}
\partial_t f_\W &= -\frac{\vp}{a^2m}\cdot \vnabla_x f_\W + mV   \frac{2}{\hbar} \sin\left(\frac{\hbar}{2}\overleftarrow{\vnabla}_{\! \!  x} \cdot\overrightarrow{\vnabla}_{\! \!  p}\right) f_\W \,.
\end{align}
The extra terms that are induced compared to the Vlasov equation are combinations of an uneven number of spatial derivatives on the gravitational potential and momentum-derivatives of the phase-space distrution $mV (\overleftarrow{\vnabla}_{\! \!  x}\cdot \overrightarrow{\vnabla}_{\! \!  p})^{2n+1} f_W$ with a prefactor of $\left(\frac{\hbar}{2}\right)^{2n} $. The $\sin$-term in \eqref{eq:WignerVlasov2} corresponds to the $\sinh$-term in the time evolution for the cumulant generator $\ln G_T$ \eqref{eq:evolutionlnGnphi} because the Fourier transform that defines the generator replaces $\vnabla_p\rightarrow i\vJ/m$.

The associated generating functional can be computed by plugging the expression for the Wigner distribution $f_\W$ in terms of $\psi = \sqrt n \exp\left(i\phi/\hbar\right)$ in \eqref{genfun}
\cite{Uhlemann14} 
\begin{align}
\label{eq:mgfshifts}
G_W(\vJ) &=\sqrt{n\left(\vx+\tfrac{\hbar}{2m}\v{J}\right)n\left(\vx-\tfrac{\hbar}{2m}\v{J}\right)}\ \exp\left\{\tfrac{im}{\hbar}\left[\phi\left(\vx-\tfrac{\hbar}{2m}\v{J}\right)-\phi\left(\vx+\tfrac{\hbar}{2m}\v{J}\right)\right]\right\} \,,\\
\label{eq:cgfshifts}
\ln G_W(\vJ) &= \frac{1}{2} \left[\ln n\left(\vx+\tfrac{\hbar}{2}\v{J}\right)+ \ln n\left(\vx-\tfrac{\hbar}{2}\v{J}\right)\right]\ + \frac{im}{\hbar}\left[\phi\left(\vx-\tfrac{\hbar}{2}\v{J}\right)-\phi\left(\vx+\tfrac{\hbar}{2}\v{J}\right)\right] \,.
\end{align}
By using that a shift in the argument of a function $f$ can be expressed through an exponential derivative operator acting on the same function $f(\vx\pm\frac{\hbar}{2m}\vJ)=\exp(\pm\frac{\hbar}{2m}\vJ\cdot \vnabla)f(\vx)$, one can rewrite the Wigner generating functional $\ln G_W$ as the solution from the finitely generated cumulants ansatz $\ln G_T$ \eqref{eq:cgfcoshsinh} with $\tilde c=\hbar/(2m)$ and recognise the associated cumulant recursion that was presented in the previous subsection. 

Hence, we have established that the Wigner ansatz for the phase-space distribution $f_W$~\eqref{eq:Wignerf} is in fact an ansatz of finitely generated cumulants of the special form $\ln G_T$ \eqref{eq:cgfcoshsinh}. Note that, for the purpose of approximating Vlasov-Poisson dynamics, $\tilde c=\hbar/(2m)$ can be regarded as free parameter that simultaneously controls the proximity to Vlasov-Poisson, and the level of restriction for the ansatz of the phase-space distribution. While the formal classical limit $\hbar \rightarrow 0$ at the level of the Wigner distribution~\eqref{eq:Wignerf} reduces to the perfect pressureless fluid model \eqref{eq:fdust} where all higher-order cumulants vanish, one can rely on a (weak) semi-classical limit where one evolves the Schr\"odinger-Poisson equation for an infinitesimally small, but nonzero $\hbar$ and then averages over small phase-space patches of size $\hbar$, as explained next.

\subsubsection{The quantum-classical correspondence for coarse-grained dynamics}

In the context of quantum mechanics, it is a well known fact that the Wigner distribution~\eqref{eq:Wignerf} is not positive semidefinite, but can go negative. This behaviour is confined to small phase-space regions on which quantum effects cause fast oscillations. If one performs a Gaussian smoothing of the Wigner distribution with spatial and momentum filter scales $(\sigma_x,\sigma_p)$ is of appropriate size for a semi-classical description
\begin{align}
\label{eq:uncertainty}
\sigma_x\sigma_p \geq \hbar/2\,,
\end{align}
one obtains a new distribution, called coarse-grained Wigner or Husimi distribution \cite{Husimi40}, which is guaranteed to be non-negative \cite{Cartwright75}. This procedure is not just a mathematical trick to obtain a sensible phase-space distribution, but also facilitates a comparison between quantum and classical physics in phase-space. When coarse-grained on scales that satisfy~\eqref{eq:uncertainty}, the time evolution of the phase-space distributions with underlying quantum and classical dynamics indeed are in close correspondence \cite{Takahashi89} and one can approach the semi-classical limit $\hbar\rightarrow 0$ in a mathematically sound way \cite{Lions93,Gerard97,Mauser02,AgisMauser08,Agis15}.

\section{Conclusion: Lessons learned and ways forward}
\label{sec:Conclusion}

\paragraph{Summary.}
In this work, we addressed the question of how to deal with the infinite cumulant hierarchy that is induced by the nonlinear gravitational dynamics of collisionless dark matter, described by the Vlasov-Poisson equation. We argued that generalising the idea of finitely generated cumulants, put forward for probability distribution functions \cite{PistoneWynn99}, to phase-space distribution functions offers a way to tackle cumulant hierarchies beyond truncation. We demonstrated the usefulness of this idea by adopting a simple ansatz for the cumulant generator \eqref{eq:lnGoperatornphi}, relying on cold initial conditions, and deriving an approximate closure scheme~\eqref{eq:cgfcoshsinh}. We showed that this constitutes a \textit{construction} of the Schr\"odinger-Poisson system as approximate dynamics for Vlasov-Poisson, thus complementing the common inverse inference of Vlasov-Poisson as semi-classical limit of Schr\"odinger-Poisson.

The idea of finitely generated cumulants could pave the way for deducting a whole class of approximate schemes for the nonlinear gravitational dynamics in Vlasov-Poisson. In particular, it can be useful to address more general initial conditions that cannot be incorporated in two scalar degrees of freedom, such as neutrinos with initial velocity dispersion.

\paragraph{Wide scope across physics.} Throughout our deduction, we did not use the precise form of the Poisson equation that determines the (gravitational) potential $V$ entering the Vlasov equation. Hence, we expect that our procedure of deriving approximate closed-form equations for cumulants will be applicable to a larger class of potentials, in particular electromagnetic potentials relevant in plasma physics. The only property we had to assume was that the potential does not vary on very small scales which should be warranted for a wide range of mean field approaches.

 In a broader context, solving nonlinear collisionless equations of motion is a common problem in many areas of physics. Hence, the idea of finitely generated cumulants could be useful for finding approximate techniques to tackle infinite cumulant hierarchies in the nonlinear regime and devise closure strategies beyond truncation. In the context of Vlasov-Poisson, we have seen that the idea of finitely generated cumulants is intimately related to a quantum version of Vlasov-Poisson. This suggests that one should take full advantage of quantum-classical correspondence by not only using semi-classical methods to understand quantum dynamics \cite{Haas05}, but also quantal methods to tackle classical dynamics \cite{Bertrand80,WK93,Uhlemann14}. In particular, it would be interesting to study to which extent those ideas can address plasma physics phenomena \cite{Davidson72plasma} or turbulence \cite{Batchelor53,Frisch95}.

\section*{Acknowledgements}
CU kindly acknowledges funding by the STFC grant RG84196 `Revealing the Structure of the Universe'. CU also wishes to thank the Wolfgang Pauli Institute for hospitality during the refinement of this work and the participants of the workshops on `Vlasov-Poisson in cosmology and plasma physics: monokinetic and multi- beam/stream solutions' and `Numerics in cosmology: around the Schr\"odinger method' for interesting and insightful discussions. Special thanks go to Cornelius Rampf for his diligent proofreading and constructive criticism of the manuscript.

\newcommand{\apj}{Astrophys. J.}
\newcommand{\apjl}{Astrophys. J. Letters}
\newcommand{\apjs}{Astrophys. J. Suppl. Ser.}
\newcommand{\mnras}{Mon. Not. R. Astron. Soc.}
\newcommand{\pasj}{Publ. Astron. Soc. Japan}
\newcommand{\apss}{Astrophys. Space Sci.}
\newcommand{\aap}{Astron. Astrophys.}
\newcommand{\physrep}{Phys. Rep.}
\newcommand{\mpla}{Mod. Phys. Lett. A}
\newcommand{\jcap}{J. Cosmol. Astropart. Phys.}
\newcommand{\prl}{Phys. Rev. Lett.}
\newcommand{\prd}{Phys. Rev. D}
\newcommand{\nat}{Nature}

\bibliographystyle{unsrt}
\bibliography{HusimiVlasovbib}

\appendix

\end{document}